\documentclass[twocolumn,preprintnumbers,amsmath,amssymb]{revtex4-1}
\usepackage{graphicx}
\usepackage{dcolumn}
\usepackage{bm}
\newcommand{\bfr}{\mathbf{r}}

\begin{document}

\title{Synthetic charge-flux quantum liquids}
\author{Tapio P. Simula}
\affiliation{School of Physics, Monash University, Victoria 3800,
Australia} 

\begin{abstract}
We apply rotating optical flux lattices to spinor Bose-Einstein condensates. Distinct quantum states emerge for fractional ratios of vortex charge density to optical flux density. We exhibit the calculated charge-flux states and discuss their topological structure and experimental signatures.
\end{abstract}
\maketitle

In the presence of a magnetic field $\bf B$, an electric charge of strength $e$ moving at a velocity ${\bf v}$ experiences a Lorentz force ${\bf F}=e{\bf v}\times {\bf B}$, which causes it to travel in a circular trajectory in the plane perpendicular to the magnetic field. In strong fields electrons can become trapped by the magnetic flux. Quantum mechanically an external magnetic field ${\bf B}=\nabla\times {\bf A}$ couples to the motion of the electrons via the magnetic vector potential ${\bf A}$ corresponding to the Hamiltonian $H=({\bf p} -e{\bf A})^2/2m$, where ${\bf p}$ is the momentum of an electron with mass $m$. Integer and fractional quantum-Hall effects in a two-dimensional electron gas emerge when the number of electric charges per magnetic flux quantum, becomes sufficiently small. Typically such conditions can be achieved by subjecting the sample to a very strong external magnetic field \cite{Klitzing1980a,Paalanen1982a,Tsui1982a,Laughlin1982a}.

In scalar superfluids, circulation of the superfluid velocity is quantized in integer units $\kappa=h/m$ given by the Onsager-Feynman rule for quantization of circulation $\oint{\bf v}\cdot {\bf dl}=n \kappa$, where $h$ is Planck constant and $n$ is an integer \cite{Leggett2006a}. This furnishes a topological invariant $\kappa$ in the system. When vortices move relative to the flow field they are immersed in, they experience a Magnus force, which is the hydrodynamic counterpart to the magnetic Lorentz force. Within the peculiar analogy between electric charges and vortices in hydrodynamics, dating back to the works of Helmholtz and Kirchoff, the quantized vortices can, in two spatial dimensions, be viewed as the analogs of electric charges in a logarithmically interacting screened two-dimensional Coulomb gas \cite{Sonin1987a,Minnhagen1987a}.  

The connection between superfluid vortices and electric charges motivates the search for analogue electromagnetic phenomena in neutral superfluids. Indeed, if the fluid is made to rotate at an angular speed $\Omega$, the Hamiltonian, when expressed in the rotating frame, acquires an extra term $-{\bf \Omega}\cdot \bfr\times{\bf p}$ due to the angular momentum of the particles. The Hamiltonian $H={\bf p}^2/2m -{\bf \Omega}\cdot \bfr\times{\bf p}$ may be re-expressed in the form $H=({\bf p}_v -  \kappa {\bf A}_\Omega)^2/2m_v -m\Omega^2 r_\perp^2/2$, where $m_v=\gamma m$ and ${\bf p}_v=\sqrt{\gamma}m$ are the effective mass and momentum of a vortex defined by the parameter $\gamma$, and ${\bf A}_\Omega=\frac{\sqrt{\gamma}}{\kappa} m\Omega( - y{\bf e}_x + x{\bf e}_y )$ is a vector potential. The effect of rotation $\Omega$ on a neutral atom thus appears equivalent to the effect experienced by a vortex charge $\kappa$ in a combination of a magnetic field ${\bf B}_\Omega =\nabla\times {\bf A}_\Omega$ and a scalar potential of the centrifugal kind. 

To emulate large effective magnetic fields ${\bf B}_\Omega$, neutral superfluids must be rotated at high angular speeds exerting a large centrifugal effect on the superfluid \cite{Abo-Shaeer2001a,Bretin2004a,Schweikhard2004a,Williams2010a,Cooper2008a,Anderson2010a}. Recently, the creation of synthetic gauge fields utilizing the spin-degrees of freedom of neutral atoms \cite{Lin2009a,Lin2011a,Lin2011b,Dalibard2011a} has opened up the possibility of generating pure synthetic magnetic fields, thereby avoiding the undesired effects of the centrifugal scalar potential. Such synthetic gauge fields may eventually facilitate the approach to a regime of strong atom-atom correlations where the vortex density becomes comparable to the atom density, possibly allowing the emergence of novel fractional quantum Hall-like states of neutral atoms. 

In this paper we investigate the effects of rotating a spinor Bose-Einstein condensate using a recently introduced optical flux lattice \cite{Cooper2011a,Cooper2011b}. We study the regime where the atoms are weakly interacting and their number is much greater than the number of vortices in the system. We show that new quantum states emerge in this system when the ratio of vortex density to the optical flux density acquires rational values. Experimentally, such regime is achievable in slowly rotating and optically thick atomic clouds.

We model a gas of ultra-cold atoms cooled below the critical temperature for Bose-Einstein condensation in the $S=1$ hyperfine-spin manifold confined in a parabolic potential well $V_{\rm trap}(\bfr)=m\omega_\perp^2 (x^2 + y^2+\lambda_z z^2)/2$, where $\omega_\perp$ is the transverse harmonic trap frequency and $\lambda_z$ is the aspect ratio. In the presence of a field ${\bf B}_\phi$, which couples to the spin ${\bf S}$ and rotates at an angular velocity $\bf \Omega$, we model this system with a Hamiltonian
\begin{eqnarray}
 &&H= -\hbar^2\nabla^2 /2m + V_{\rm trap}(\bfr) -\mu_\alpha - {\bf S}\cdot {\bf B}_\phi -{\bf \Omega}\cdot {\bf L}  +\nonumber \\
 &&gn(\bfr) + g_s \sum_\sigma M_\sigma(\bfr) S_\sigma, 
\label{eq8}
\end{eqnarray}
where the total particle density $n(\bfr)=\sum\psi^*_\alpha(\bfr)\psi_\alpha(\bfr)$ is expressed in terms a three-component spinor order parameter $\Psi = [\psi_\uparrow,\psi_0,\psi_\downarrow]^T$ and $\mu_\alpha$ are the corresponding chemical potentials and $\bf L$ is the orbital angular momentum operator. The diagonal operators are expressed as scalars. The Cartesian components of the magnetization density $M_\sigma(\bfr)=\Psi^\dagger(\bfr) S_\sigma \Psi(\bfr)$ are obtained as spin-space expectation values of the components of the spin-1 operator. Depending on the form and origin of ${\bf B}_\phi$, the term $-{\bf S}\cdot {\bf B}_\phi$ may account for an internal spin-orbit coupling or a coupling of a magnetic moment to an external field. The repulsive $g>0$ particle interactions are modeled using the standard contact interaction potential \cite{Leggett2006a} and the spin-exchange coupling $g_s$ of the atoms is set to be ferromagnetic. The parameters used in our calculations are $\lambda_z = 2$, $gN/\hbar \omega_\perp a_0^3=2000$, where $N$ is the number of atoms in the system, $g_s = -0.01g$, and $a_0=\sqrt{\hbar/m\omega_\perp}$. The chemical potentials are $\mu_\uparrow=\mu_\downarrow\approx 12\hbar\omega_\perp$ and $\mu_0=\mu_\downarrow+100\hbar\omega_\perp$. For the results presented we have chosen the scaling $a_0=1 \mu$m.

Inspired by the optical flux lattice scheme introduced by Cooper and Dalibard \cite{Cooper2011a,Cooper2011b}, we utilize an optical flux lattice of the form  
\begin{equation}
{\bf B}_\phi=g_\phi (\cos(kx){\bf e}_x+ \cos(ky){\bf e}_y +\sin(kx)\sin(ky) {\bf e}_z),
\end{equation}
where $k$ is a wavenumber and $g_\phi=3\hbar\omega_\perp$ determines the coupling strength of the optical flux lattice which has an areal flux density $\mathcal{B}_\phi=\ell_\phi (k/\pi)^2$. We set the flux lattice rotating at an angular speed $\Omega$, which controls the vortex charge density nucleated in the system. The key point here is that it is possible to control $\bf \Omega$ and ${\bf B}_\phi$ and hence the number of vortex charges and the number of optical flux quanta independently of each other. We have shifted the origin of the optical flux lattice by $\pi/2$ to set a centre of a flux cell on the axis of rotation. That the superposed rotating lattice carries optical flux quanta in a form of optical vortices, imprinting geometric phases on the condensate, is the crucial distinguishing factor between the optical flux lattice scheme \cite{Cooper2011a,Cooper2011b} and the conventional, non-topological, optical lattices which have previously been applied to nucleate vortex charges in Bose-Einstein condensates \cite{Tung2004a,Williams2010a}. Dynamical variation of the lattice spacing by a factor of 2.5 has also been realized for conventional optical lattices \cite{Foot2010a}. Similar experimental techniques could also be deployed to rotate optical flux lattices and to control their flux density.

We define a charge-flux ratio $\nu=p/q$, where $p$ and $q$ are the areal density of vortex circulation quanta $p=\ell_vN_v/A$ and the optical flux quanta $q=\ell_\phi N_\phi/A$, respectively. In our system the charges are half-quantum vortices $\ell_v=\frac{1}{2}$ and each flux lattice cell encloses half a flux quantum $\ell_\phi=\frac{1}{2}$ \cite{Cooper2011a,Cooper2011b}, such that $\nu=N_v/N_\phi$. We extract $\nu$ from the calculated states by counting the number of vortex charges $N_v$ inside an area containing $N_\phi$ flux lattice cells. 

Figure 1 (a) exhibits a calculated column density $n_z(x,y)$ of the condensate integrated along the $z$-axis when an optical flux lattice is applied in a charge neutral (nonrotating) system. The corresponding magnetization column density $|M_z(x,y)|$ is shown in Fig.~1 (b), which reflects the square lattice structure of the underlying optical flux lattice. Figure 1 (c) and (d), respectively, depict the condensate and magnetization column densities  for a synthetically charged (rotating) system in the absence of the external optical flux. Circles in Fig.~1 (c) mark the locations of the vortex charges whose long-range repulsive vortex-vortex interactions results in their crystalline ordering. These vortex charges are half-quantum vortices and due to their composite structure, the total atom density remains nearly uniform inside the vortex cores.

% FIGURE ======================
\begin{figure}
\includegraphics[width=0.9\columnwidth]{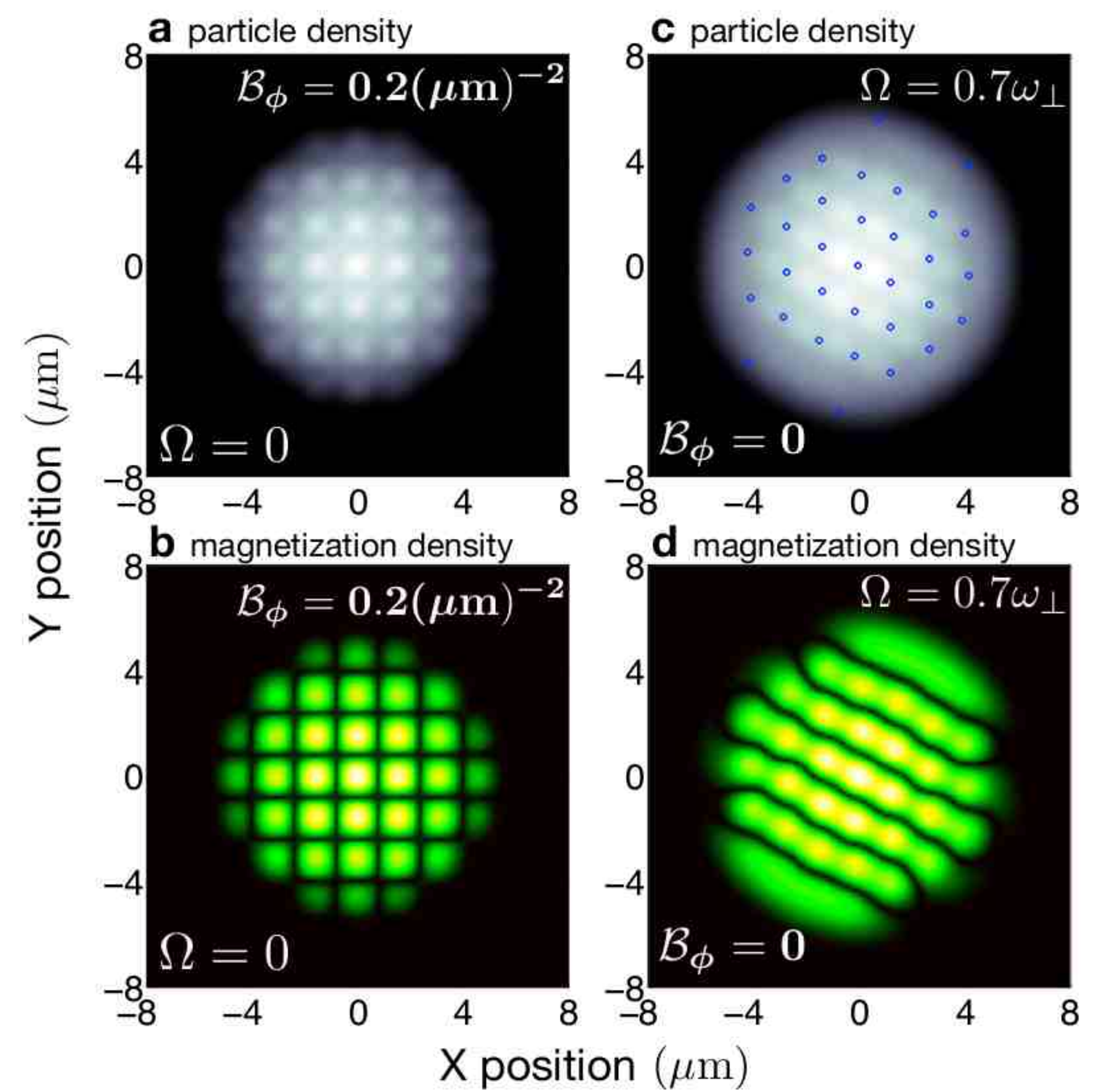}
\caption{(Color online) Flux-only and charge-only states. Condensate column density (a) and magnetization density (b) for a charge neutral system in the presence of an optical flux. Condensate column density (c) and magnetization density (d) for a synthetically charged system in the absence of optical flux. The circles in (c) mark the locations of the half-quantum vortex charges. The rotating condensate (c) has a larger diameter compared to that of the nonrotating cloud (a) due to the centrifugal effect.}
\label{fig1}
\end{figure}
% FIGURE ======================

% FIGURE ======================
\begin{figure}
\includegraphics[width=1\columnwidth]{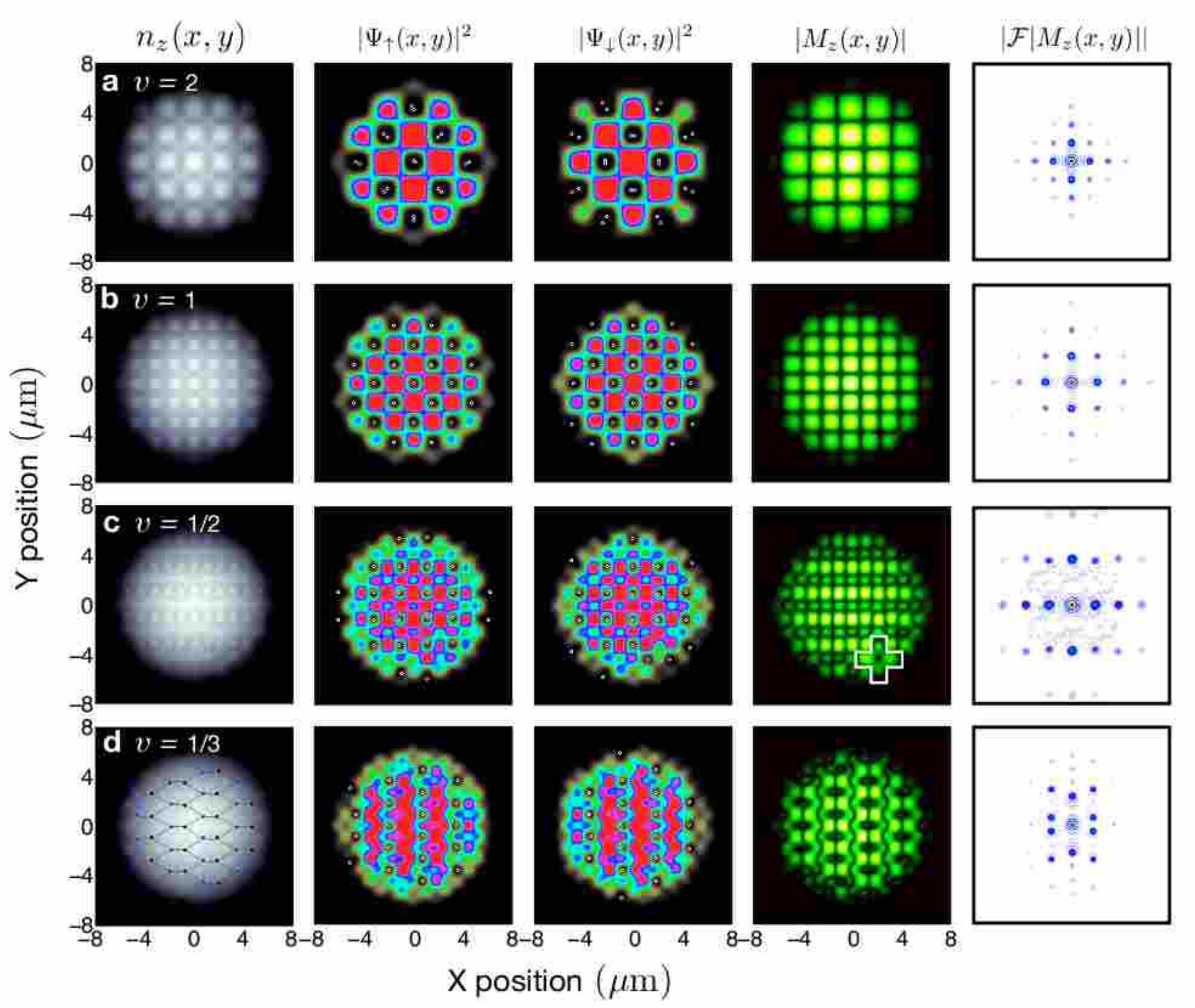}
\caption{(Color online) Synthetic charge-flux states with different fractional values of $\nu$. Condensate column density (first column) and corresponding spin-up (second column) and spin-down (third column) component densities and magnetization column density (third column) and its Fourier transformation (last column) corresponding to different ratios $\nu$ marked on each row (a)-(d). Circles in second and third column frames mark the locations of the half-quantum vortex charges. Two Frenkel-type defects are highlighted by a cross in the fourth frame of row (c). Optical flux densities in (a) - (d) are $\mathcal{B}_\phi=\{0.10, 0.20, 0.46$ and $0.62\}$ ($\mu$m)$^{-2}$, respectively. The optical flux lattice rotation frequency is $\Omega = 0.7 \omega_\perp$ for all of these states.}
\label{fig1}
\end{figure}
% FIGURE ======================

When the optical flux lattice is combined with finite vortex charge density, charge-flux correlations develop in the system resulting in quantum phases which feature self-assembled topological structures. Charge-flux commensurability is achieved when the ratio of the vortex charge density $n_v$, which may be estimated by the Feynman's rule $n_v=2\Omega/\kappa$ to the optical flux density $\mathcal{B}_\phi$ is a rational number $1/i=n_v/\mathcal{B}_\phi$. For our system the corresponding commensurable wavenumbers are $k_i=2\pi\sqrt{i \Omega /\kappa}$. On varying $k$ and/or $\Omega$, commensuration energy related to inter-vortex interaction and topological pinning of vortex charges by the optical flux quanta is anticipated to drive the system into new ground states. Effects of commensurability between a vortex lattice and a non-topological optical lattice have been studied previously in both weakly interacting and strongly correlated systems and the results are reviewed in \cite{Cooper2008a}. 

Figure 2 shows numerically calculated states with different values of $\nu$ obtained for fixed $\Omega=0.7\omega_\perp$ and by varying the optical flux density $\mathcal{B}_\phi$. We choose to fix the value of $\Omega$ since the flux density is proportional to the square of $k$ whereas the vortex density is linear in $\Omega$. Thereby we also avoid having to deal with the large variation in condensate and vortex densities, which are strongly dependent on $\Omega$. The first column of Fig. 2 shows an integrated column density $n_z(x,y)$ of the condensate calculated for different optical flux densities. The second and third columns, respectively, display the density of condensate atoms in two of the hyperfine substates, which have practically equal populations. The third hyperfine state has been lifted in energy with respect to the other two states and is not shown due to its negligible population. Experimentally, this could be achieved using a quadratic Zeeman effect. The third column shows the magnetization column density $|M_z(x,y)|$ and the last row displays its Fourier transform. In Fig. 2 (a) each cell of the flux lattice traps two half-quantum vortices realizing a $\nu=2$ state while in the $\nu=1$ state (b) every flux lattice site is occupied by a single half-quantum vortex charge. Generically states with integer $\nu$ are composed of interlaced lattices of $\nu$-fold quantized vortices. In the half-filling state (c) every second row of the flux lattice is free of vortex charges. The topology of the fractional $\nu=1/3$ state is shown in the last row (d). 

By inspecting the states shown in Fig. 2 we extract a trial spinor $\Psi(\bfr) = f(\bfr)[\psi_\alpha(\bfr), \psi_\beta(\bfr)]^T$, where $T$ denotes a transpose and the function $f(\bfr)$ is a smooth envelope determining the finite system size. Here the spinor components are $\psi_\sigma(\bfr) =\prod_v (z-z_v)^{p_v} \prod_j e^{-|z_j|^2/(2\lambda^2)}$, where $j$ and $v$ respectively enumerate the flux cells and the cells occupied by charges in the state $\sigma$. For $\nu<1, p_v=1$ and $p_v=\nu$ for integer values of $\nu$ . The length $\lambda$ is a variational parameter and the index $v$ labels the complex coordinates $z_v=x_v+iy_v$ of the vortex charges. 

These quantum liquids support macroscopic topological fractional-vortex excitations which emerge as impurities such as interstitials and vacancies. A pair of such Frenkel-type defects are highlighted in Fig.~2 (c) where the two half-quantum vortices shown appear rotated 90 degrees about their centre-of-mass---half way through exchanging their equilibrium places. Experimentally, controlled vortex braiding comprising of adiabatically moving the vortices \cite{Virtanen2001a} around each other would require addressability of the optical flux lattice at a single site level. In addition to the macroscopic topological defects, these quantum liquids possess a low energy spectrum of microscopic Bogoliubov-de Gennes collective quasiparticle modes. Of particular interest is the level structure of those quasiparticle modes associated with the optical flux quanta, vortex motion and the edge localized high angular momentum surfon states. 

% FIGURE ======================
\begin{figure}
\includegraphics[width=0.6\columnwidth]{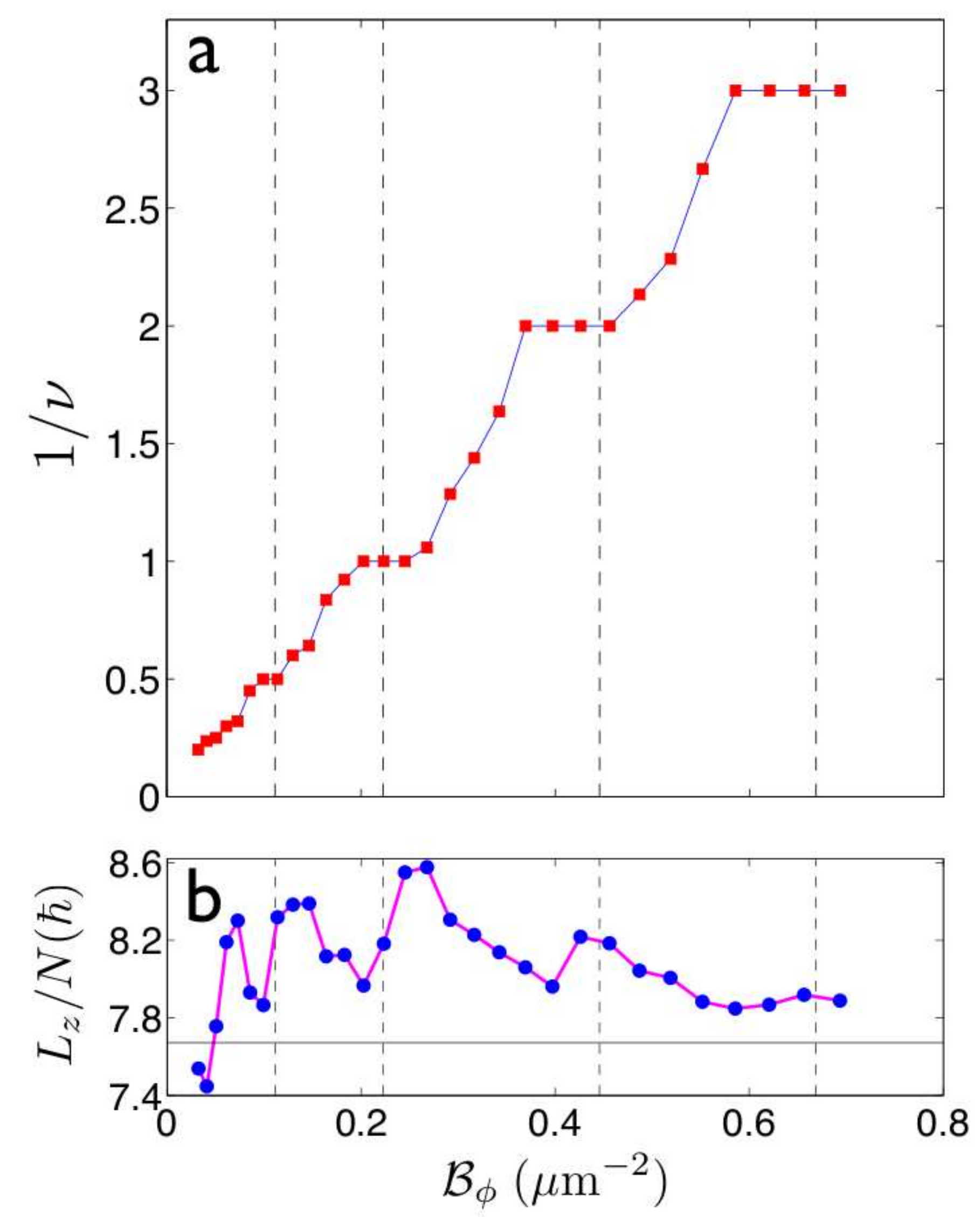}
\caption{(Color online) Ratio $1/\nu$ (a) and an orbital angular momentum per particle $L_z/N$ as functions of the optical flux density $\mathcal{B}_\phi$. The calculated data points are joined by solid lines. The vertical dashed lines are plotted at commensurate flux densities $\mathcal{B}_\phi$ with $k_i=2\pi\sqrt{i \Omega /\kappa}$ where $i$ = 1/2, 1, 2, and 3. The horizontal base line in (b) is drawn at the value corresponding to the angular momentum in the absence of the optical flux lattice shown in Fig.~1}
\label{fig1}
\end{figure}
% FIGURE ======================

% Figure 3
Figure 3 (a) shows a measurement of the ratio of the optical flux density to the vortex density $1/\nu$ as a function of the optical flux density $\mathcal{B}_\phi$. Figure.~3 (b) shows a measurement of the expectation value of the $z$-component of the orbital angular momentum per particle $L_z/N$ as a function of the optical flux density $\mathcal{B}_\phi$. The vertical dashed lines in both subfigures indicate the commensurate flux densities estimated using the Feynman's rule for the average vortex density. The commensurate wavenumbers $k_i$ seem to be slightly overestimated by this method indicating that the number of vortices in this system is not large enough for it to accurately mimic solid body rotation. The Feynman's rule has also been observed to overestimate the vortex number in experiments \cite{Williams2010a}. For commensurable ratios of $1/\nu$ we observe minima in Fig.~3(b) coincident with the plateaux in Fig.~3(a). The horizontal line in Fig.~3(b) corresponds to the angular momentum of the flux-free state shown in Fig.~1.

The obtained results can be understood in terms of interaction of the vortex charges with the optical flux quanta. At commensuration, the ratio of the average vortex charge density to the average optical flux density is a rational number, corresponding to the plateaux in Fig.~3(a)). The optical flux lattice is then rotating at an optimal angular speed with respect to its flux density such that the condensate superflow and the transport of vortices are in balance. Increasing the optical flux density introduces compressive stress to the vortex liquid forcing it to increase its density, which is tantamount to creating excitations and radial vortex currents. Higher vortex density manifests itself as an increase in the orbital angular momentum. 

Since the angular speed of the vortices is restricted by the fixed rotation frequency of the flux lattice, the superflow can not be in balance with the flow of the vortices for incommensurate charge-flux ratios. The relative motion between vortices and condensate flow gives rise to mutual friction \cite{HallVinen} in the presence of excitations. The vortices resist compression by creating vortex currents which restore the balance between vortex density and flux density. This facilitates transitions between quantum states with different lattice symmetries shown in Fig.~2. 

The staircase behavior seen in Fig.~3(a) and the corresponding angular momentum fluctuations shown in Fig.~3(b) result from the alternating compression and relaxation of the vortex charge density. The compressibility of the background gas of condensate atoms due to the gapless sound wave spectrum does not exclude the possibility of coexistence of an incompressible vortex liquid since the part of the excitation spectrum which couples to the motion of vortices (Kelvin-Tkachenko modes) may be gapped and decoupled from the compressional sound waves.

The predicted charge-flux states shown in Fig.~2. can be prepared experimentally by applying an optical flux lattice \cite{Cooper2011a,Cooper2011b} to spinor Bose-Einstein condensates and rotating it at a fixed angular speed. Such rotating drive has previously been achieved using non-topological optical lattices \cite{Tung2004a,Williams2010a}. Transitions between different charge-flux states can be observed by varying the wavenumber of the optical flux lattice \cite{Foot2010a}. Subsequently the condensate density can be imaged using standard absorption or phase-contrast imaging techniques, either in-situ or after a time-of-flight and the different spin-components can be resolved individually by separating them using Stern-Gerlach fields. The ratio $\nu$ of vortex charge density to optical flux density can be extracted from the obtained images. The magnetization structure which clearly reveals the distinct lattice symmetries of the charge-flux states could be directly measured using magnetization-sensitive phase contrast imaging \cite{Higbie2005a,Vengalattore2008a}. The orbital angular momentum can be inferred from spectroscopic measurements of the collective excitation spectrum of the condensate \cite{Chevy2000a}. Resonance spectroscopy can also be used to explore the low-lying excitation spectrum of these quantum liquids.

In conclusion, we have computationally applied a rotating optical flux lattice to spinor Bose-Einstein condensates and have found distinct quantum states and ordering of the nucleated fractional-charge vortices at rational ratios of the number of vortex charges to the number of optical flux quantum. These charge-flux states could be observed in slowly rotating and optically thick cold atom clouds using current experimental technology. It will be particularly interesting to probe the excitation spectra to reveal the gapped states and zero modes of these quantum liquids.

% References ==========================

\end{document}